\documentclass[referee,a4paper,12pt,traditabstract]{swsc} 

\usepackage{graphicx}
\usepackage{txfonts}
\usepackage{subfigure}
\usepackage{epstopdf}
\usepackage[displaymath,mathlines]{lineno}
\usepackage[authoryear,round]{natbib}
\usepackage[backref]{hyperref}
\usepackage{url}

\usepackage{lscape}

\hypersetup{colorlinks=true,citecolor=cyan,urlcolor=cyan,linkcolor=blue}

\begin{document}


   \title{Synoptic solar observations of the Solar Flare Telescope focusing on space weather}

   
   \titlerunning{Synoptic solar observations of the SFT}

   \authorrunning{Hanaoka et al.}

   \author{Yoichiro Hanaoka \inst{1}
          \and              Takashi Sakurai \inst{1}
          \and              Ken'ichi Otsuji \inst{2}
          \and              Isao Suzuki \inst{3}
          \and             Satoshi Morita\inst{1}
          }

   \institute{National astronomical Observatory of Japan,
              2-21-1 Osawa, Mitaka, Tokyo181-8588, Japan\\
              \email{\href{mailto:yoichiro.hanaoka@nao.ac.jp}{yoichiro.hanaoka@nao.ac.jp}}
         \and
             National Institute of Information and Communications Technology\\
         \and
             Bunkyo Gakuin University\\
             }


 
  \abstract
   {The solar group at the National Astronomical Observatory of Japan is conducting synoptic solar observations with the Solar Flare Telescope. While it is a part of a long-term solar monitoring,  contributing to the study of solar dynamo governing solar activity cycles, it is also an attempt at contributing to space weather research.
The observations include imaging with filters for H$\alpha$, Ca K, G-band, and continuum, and spectropolarimetry at the wavelength bands including the He I 1083.0 nm / Si I 1082.7 nm and the Fe I 1564.8 nm lines. Data for the brightness, Doppler signal, and magnetic field information of the photosphere and the chromosphere are obtained.
In addition to monitoring dynamic phenomena like flares and filament eruptions, we can track the evolution of the magnetic fields that drive them on the basis of these data. Furthermore, the magnetic field in solar filaments, which develops into a part of the interplanetary magnetic cloud after their eruption and occasionally hits the Earth, can be inferred in its pre-eruption configuration. 
Such observations beyond mere classical monitoring of the Sun will hereafter become crucially important from the viewpoint of the prediction of space weather phenomena. The current synoptic observations with the Solar Flare Telescope is considered to be a pioneering one for future synoptic observations of the Sun with advanced instruments.
   }        

\keywords{Sun -- synoptic observation -- instrumentation -- eruptive events -- magnetic field}

   \maketitle

\section{Introduction}

Solar observations covering the full-disk of the Sun, namely the ``synoptic'' ones, have long been demanded from the viewpoint of space weather \citep[e.g.,][]{2019BAAS...51c.110M}. Besides observations to detect the cyclic variation of solar activity such as sunspot counting, those aiming to detect short-lived dynamic phenomena like flares and eruptive events occurring at any place on the Sun are also necessary. One of the representative observations of this kind is flare monitoring using the H$\alpha$ line, a prominent chromospheric line of H I 656.3 nm. Currently, there are some active world-wide networks for H$\alpha$ synoptic observations (Global Oscillation Network Group [GONG], \cite{2018SpWea..16.1488H}; Global High Resolution H-alpha Network, \cite{2000ESASP.463..617S}).

In recent times, the influence of solar phenomena on our environment has increased, and the demand for observational data of various aspects of the Sun is increasing. 
The monitoring observations of dynamic phenomena on the solar surface have become more advanced, and at present, X-ray, extreme ultraviolet, and coronagraphic observations from space provide solar monitor data all the time. However, they are basically imaging observations to record structural changes projected onto the sky plane seen from the Earth, with some exceptions. On the other hand, some ground-based observatories are obtaining Doppler velocity data in H$\alpha$ \citep[e.g.,][]{2017SoPh..292...63I}. The combination of the two-dimensional images and Doppler data gives the three-dimensional velocities of eruptive features. In the case of filament eruptions in particular, which often develop into coronal mass ejections (CMEs), the three-dimensional velocity information is essential for determining the initial mass motion of CME plasma. H$\alpha$ observations are the most suitable for obtaining such data. 

In addition, the prediction of flares and eruptive events is gaining importance, because it allows us to prepare and avoid or mitigate their harmful effects. Tracking the evolution of the photospheric vector magnetic field inferred from polarization measurements of the photospheric absorption lines is the key to monitor the magnetic energy storage which leads to the occurrences of flares and eruptions \citep[e.g.,][]{2012ApJ...760...31K}. 
Furthermore, the measurement of the chromospheric magnetic field is gaining attention, owing to its proximity to the site where the magnetic activity actually occurs rather than the photosphere, as shown by the polarization measurements of cool coronal loops and flaring atmosphere \citep[e.g.,][]{2016ApJ...833....5S, 2018ApJ...860...10K, 2019ApJ...874..126K}. 
Measurements of the polarization in chromospheric absorption lines are necessary for deriving information about the magnetic field of the chromosphere.

Among the chromospheric absorption lines, the He I 1083.0 nm line has a particular advantage in the polarization measurements. In addition to the fact that its polarization signals provide information about the magnetic field of the chromosphere \citep[e.g.,][]{2007AdSpR..39.1734L}, they also show the magnetic field around the filaments \citep{2017ApJ...851..130H}. 
In the discussion about the future prospect in forecasting key CME properties  \citep{2019SpWea..17..498K}, it is considered to be valuable for space weather prediction to know the orientation of the magnetic field of the flux ropes ejected from the solar surface as a part of CMEs. The reason is that geomagnetic storms become severe upon the arrival of the southward magnetic field in flux ropes at the Earth. 
{\it In situ} measurements detect the magnetic field in flux ropes in the interplanetary space shortly before their arrival at the Earth. 
On the other hand, the polarization measurements of the He I 1083.0 nm line show the magnetic field orientation of the filaments on the solar surface, which erupt as a part of CMEs, much before their arrival. 

Regular synoptic observations of the photospheric magnetic field have been carried out for long time using both ground-based and space-borne instruments \citep[see e.g.,][]{2012LRSP....9....6M}. On the other hand, the realization of regular chromospheric magnetic field measurements has been challenging, while many {\it ad hoc} observations with large telescopes have been done successfully \citep{2014LRSP...11....2P}. The polarization signals from the chromospheric lines are weak, and therefore, require particularly low-noise polarimeters. Furthermore, at 1083.0 nm, ordinary silicon detectors such as CCDs and CMOSs have very low sensitivity, making it necessary to employ infrared detectors, which are still being developed.

Among various synoptic observation instruments, the Synoptic Optical Long-term Investigations of the Sun (SOLIS) constructed by the National Solar Observatory (NSO) of the USA is intended to realize full-disk photospheric and chromospheric imaging and polarimetry \citep{2003SPIE.4853..194K}. It has a capacity to measure the full-Stokes polarization in a photospheric line, Fe I 630.2 nm, and circular polarization for a chromospheric line, Ca II 854.2 nm. 
Imaging was available for the He I 1083.0 nm line but was subsequently replaced by the full-Stokes polarimetry at Ca II 854.2 nm \citep{2015IAUS..305..186G}.

In spite of such difficulties, we, the solar group of National Astronomical Observatory of Japan (NAOJ), have realized synoptic solar observation with the Solar Flare Telescope \citep[SFT,][]{1995PASJ...47...81S}, which enables us to image the photosphere and the chromosphere including Doppler observations and full-Stokes polarimetry to obtain magnetic field information of the photosphere, chromosphere, and filaments. NAOJ started its regular solar observation in 1917, and its observation instruments have been continuously renovated \citep{1998ASPC..140..483S, 2013JPhCS.440a2041H, 2016ASPC..504..313H}. After 2010, we completely replaced the instruments in the SFT. 
Newly installed instruments enable us to perform filter imaging observations at H$\alpha$, Ca II K (393.4 nm), G-band (430.5 nm), and green continuum along with Doppler velocity measurements in H$\alpha$, and spectropolarimetry at the wavelength bands including He I 1083.0 nm / Si I 1082.7 nm and Fe I 1564.8 nm lines. 

We adopted the H$\alpha$ line to catch the activity in the chromosphere. Some other lines such as He I 1083.0 nm provide similar information, but we were able to utilize an existing tunable narrow-band filter, which is required to realize high-cadence observations acquiring imaging and Doppler data, for the H$\alpha$ observation.
The Ca II K is also a chromospheric line. The G-band is not a single atomic absorption line, but consists mainly of many molecular lines of CH, and it is formed in the upper photosphere. The continuum provides information about the photosphere. 
The He I 1083.0 nm line provides information about the magnetic field of the chromosphere.
On the other hand, the Si I 1082.7 nm and Fe I 1564.8 nm lines are photospheric lines.
The Fe I 1564.8 nm line, which shows a particularly large Zeeman splitting, provides photospheric magnetic field data from a perspective different from that of other lines. 
The fact that these three lines can be observed with a single instrument is also an advantage of adopting near-infrared wavelengths.

Observations with the SFT are carried out daily, and 
quick-look images and polarization maps of the wavelengths mentioned above as well as real-time images and Doppler data in H$\alpha$ are available on a web page, \url{https://solarwww.mtk.nao.ac.jp/en/solarobs.html}, for the use of the space weather community as well as general public. Accumulated data including those taken with past instruments are also available in the ``database'' webpage.
In this study, we describe the instruments and provide some observational examples demonstrating the significance of such observations in space weather research.

At present, some new projects are being proposed to meet the increasing demands of solar synoptic observations focusing on space weather research. One of them is the Solar
Physics Research Integrated Network Group (SPRING) initiated mainly by the European community \citep{2018SPIE10702E..4HG}, while another is the next generation GONG (ngGONG) proposed by the NSO of the USA \citep{2019BAAS...51g..74H}. In the proposed observations, they include polarimetry in the infrared wavelengths of He I 1083.0 nm and Fe I 1564.8 nm as well as imaging and Doppler velocity measurements, like the SFT. These projects involve several telescopes with a larger aperture to be located in several sites in the world. Although the SFT is a small, single telescope, it is a pioneer for future synoptic observation instruments.

In this study, we describe the instruments of the SFT in Section 2, and present examples of observational results in Section 3. Section 4 contains the concluding remarks.

\section{Instruments}

\begin{figure}
\centering
\includegraphics[width=0.4\columnwidth]{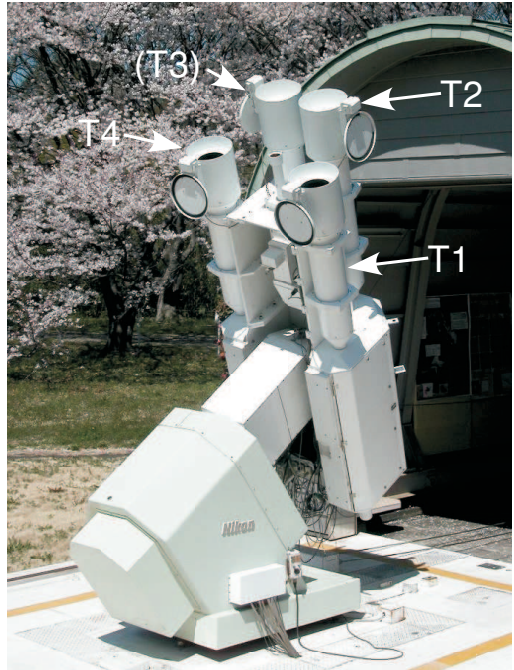}
\caption{\small Picture of the Solar Flare Telescope of NAOJ. It has four refractor tubes designated as T1--T4.
} 
\label{fig:fig1}
\end{figure}

The SFT (Fig. 1) was manufactured by Nikon and started its observations of the various aspects of solar flares in 1990 \citep{1995PASJ...47...81S}. As seen in Figure 1, it has four refractor tubes. Originally each tube had its own back-end instrument, i.e., an imaging full-Stokes polarimeter using the Fe I 630.3 nm line, photospheric Doppler imager using the Fe I 633.7 nm line, H$\alpha$ imager, and continuum imager. The imaging observations and polarimetry were performed using the field of view of 440$''\times 330''$, which is sufficient for covering an active region. From 2002, experimental polarimetry for the chromosphere in H$\alpha$ was started with a small field of view as before \citep{2006ASPC..354..324H}. Subsequently, an infrared spectropolarimeter, which observes the full-disk using the He I 1083.0 nm / Si I 1082.7 nm and Fe I 1564.8 nm lines, was installed \citep{2018PASJ...70...58S}. All the other observations were stopped to install the infrared spectropolarimeter in the SFT behind the T2 tube (see Fig. 1), and the spectropolarimeter became operational in 2010.

The infrared spectropolarimeter occupies the rear space of the two tubes, T2 and T3. In 2011, a new H$\alpha$ instrument was installed in the rear of T1 as a full-disk imager. A filter imager for G-band and continuum was installed in the rear of T4 in 2012, and in 2015, a Ca K imager was added. 

\subsection{Filter imaging instruments}

\begin{landscape}
\begin{table}
\caption{Specifications of imaging instruments installed in T1 and T4 \label{tab:table1}}
\centering
\begin{tabular}{lllll}
\hline\hline
&  \multicolumn{1}{c}{T1} &  \multicolumn{3}{c}{T4}\\
\cline{3-5}
& \multicolumn{1}{c}{H$\alpha$} & \multicolumn{1}{c}{Ca K} & \multicolumn{1}{c}{G-band} & \multicolumn{1}{c}{Continuum} \\
\hline
Objective lens & D = 125 mm f = 2250 mm &\multicolumn{3}{c}{D = 125 mm f = 2250 mm}\\
   & \quad (Nikon)   &\multicolumn{3}{c}{(Nikon)}\\
Beam splitter &  &  \multicolumn{3}{c}{50/50 cube beam splitter (Edmund Optics )}\\
Filter ($\lambda_0$/FWHM) & Prefilter 656.3 nm/2.7 nm  &   Heat Rejection 400 nm/50 nm      & 430.5 nm/1.0 nm & ND 1\% \\
        &  \quad (Andover)                    & \quad (Edmund Optics) & \quad (Materion Barr)  & G530 530 nm/50 nm (Hoya) \\
      & Lyot 656.3 nm/0.025 nm (Zeiss) & 393.4 nm/ 0.2 nm    &    ND 10\%       & BK7 spacer \\
        &     ND 50\%                     &  \quad (DayStar)      & \multicolumn{2}{c}{IR-cut HA-50 (Hoya)} \\
Field lens    & D = 40 mm f = 120 mm            & D = 40 mm f = 120 mm & \multicolumn{2}{c}{D = 40 mm f = 120 mm}             \\
Imaging lens & 50 mm F2.8 Macro (Sigma) & 50 mm F2.8 Macro (Sigma) & \multicolumn{2}{c}{50 mm F2.8 Macro (Sigma)} \\
Camera & Andor Zyla 4.2   &   Bitran BH-52L &             \multicolumn{2}{c}{Photonfocus MV1-D2080-160}  \\
Detector format & 2048$\times$2048 6.5$\times$6.5$\mu$/pixel & 2048$\times$2048 7.4$\times$7.4$\mu$/pixel & \multicolumn{2}{c}{2080$\times$2080 8$\times$8$\mu$/pixel} \\
A/D, Frame rate & 16 bit, 100 fps & 12 bit, 8 fps & \multicolumn{2}{c}{12 bit, 34 fps} \\
Spatial Sampling & 1.1$''$pixel$^{-1}$ & 1.0$''$pixel$^{-1}$ & \multicolumn{2}{c}{1.0$''$pixel$^{-1}$}  \\
Cadence & every 30 s (H$\alpha$ center)             &   every 30 s      &  every 5 min    & every 5 min \\
    & every 150 s (H$\alpha$ offband) & & & \\
Exposure & 1 ms / 4 ms / 16 ms  & 160 ms (320 ms, 640 ms)$^{\mathrm{a}}$ & 0.7 ms (1.4 ms, 2.8 ms)$^{\mathrm{b}}$ & 1.6 ms (3.2 ms, 6.4 ms)$^{\mathrm{b}}$ \\
Frame selection & every 30 images  & every 10 images & every 30 images & every 30 images \\
\hline\hline
\multicolumn{5}{l}{$^{\mathrm{a}}$ Owing to the deterioration of the filter, prolonged from 10 ms (20 ms, 40 ms) at the beginning.}\\
\multicolumn{5}{l}{$^{\mathrm{b}}$ Before the installation of the Ca K system, exposure times were half of these values.}\\
\end{tabular}
\end{table}
\end{landscape}
The H$\alpha$ imager of T1 and the Ca K, G-band, and continuum imager of T4 are filter imaging instruments. Solar images at these wavelengths are taken with 2k$\times$2k cameras with 1.0--1.1$''$ spatial samplings. Sample images collected by the filter instruments along with a polarization map taken with the infrared spectropolarimeter are shown in Figure 2. A schematic diagram of the filter imaging instruments is shown in Figure 3. The specifications of the optical elements and cameras are listed in Table \ref{tab:table1}.

\begin{figure}
\centering
\includegraphics[width=1.0\columnwidth]{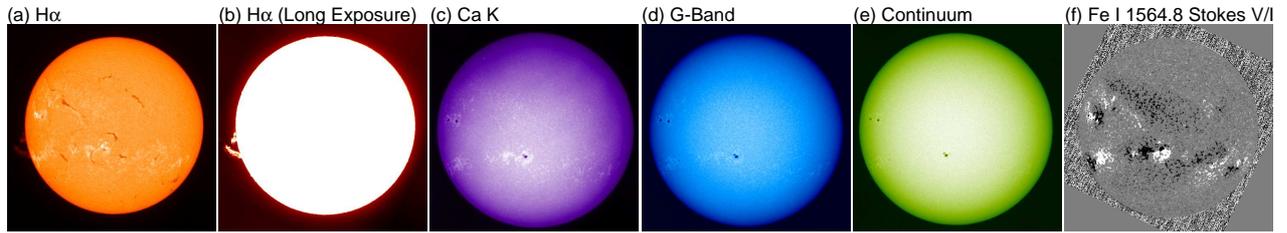}
\caption{\small Sample full-disk images of the Sun on October 19, 2015 taken using the SFT. The images in panels (a)--(b) were taken with the T1 H$\alpha$ imager while those in (c)--(e) were taken with the T4 filter imager. The circular polarization (Stokes $V/I$) map in (f) was obtained using the infrared spectropolarimeter; the display range is $\pm0.2$ \%.
} 
\label{fig:fig2}
\end{figure}

\begin{figure}
\centering
\includegraphics[width=1.0\columnwidth]{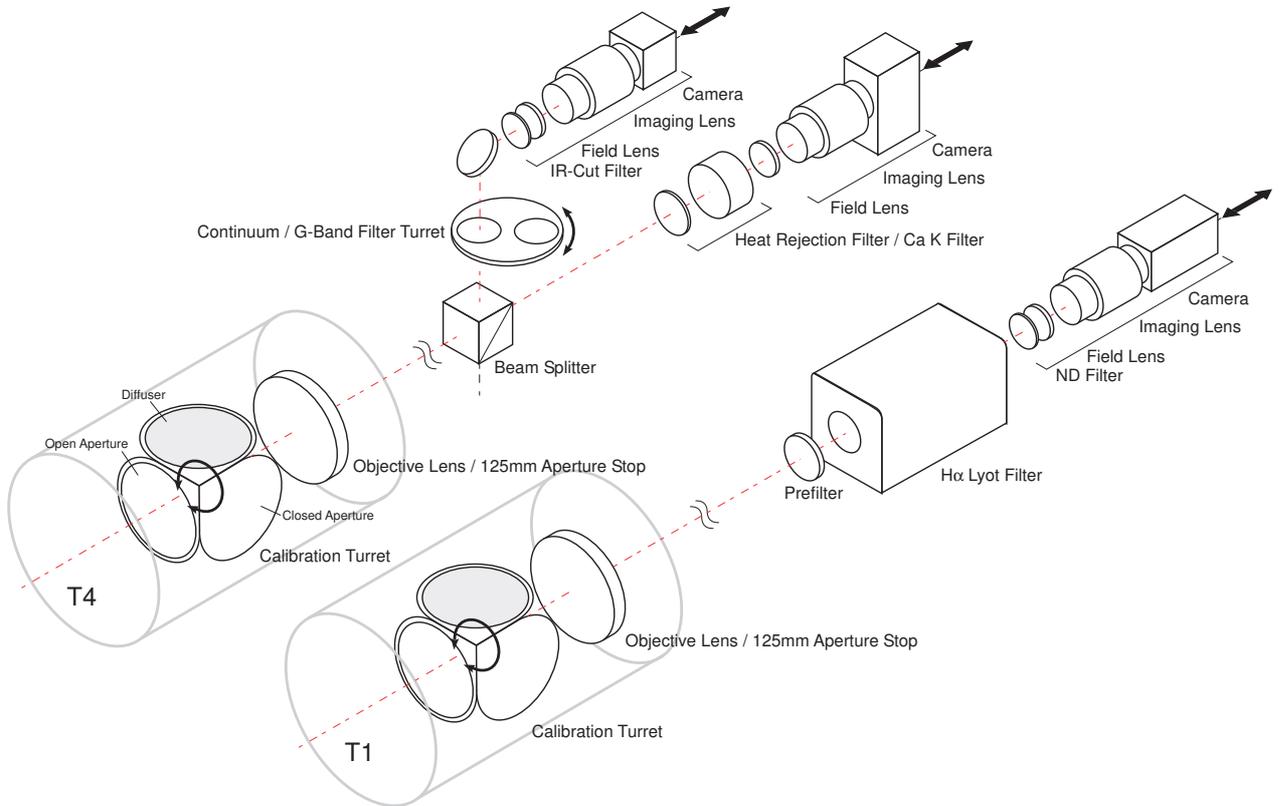}
\caption{\small Schematic diagram of the filter imager instruments installed in the rear sides of T1 (H$\alpha$) and T4 (Ca K, G-band, and continuum). Red dash-dot lines indicate the optical axes, while the arrows show motions of movable parts.
} 
\label{fig:fig3}
\end{figure}

\subsubsection{T1 H$\alpha$ observation system}

As shown in Figure 3, the H$\alpha$ observation system installed in the rear of T1, has a 150 mm achromatic objective lens used with a 125 mm aperture stop. The H$\alpha$ filter is a Zeiss Lyot filter with a passband of 0.025 nm and a tunable range of 656.3 nm$\pm$1.6 nm. The aperture stop changes the focal ratio to f/18, which is suitable for a beam incident onto the Lyot filter. The transmission wavelength of the Lyot filter is tuned by a microcomputer.

The H$\alpha$ images are recorded using a scientific-CMOS (sCMOS) camera, Andor Zyla 4.2, with a readout noise lower than 1 e$^-$. Its low noise-level enables the capture of low light-level structures such as faint prominences with a sufficiently high signal-to-noise ratio (SNR). The previous camera, Bitran BH-52L (currently used for the Ca K imaging) is an interline CCD camera. We experienced a smear effect with it, which occurs when the exposure time is much shorter than a frame time (=1/frame rate). Owing to the smear effect, a part of the photoelectrons of the solar disk image spreads out into the sky pixels, thereby deteriorating the image quality outside the limb, in particular. To prolong the exposure time with a neutral density filter is a way to reduce the smear effect, but it results in the image degradation due to the seeing effect. Therefore, we replaced the camera.

The focus position of the camera gradually changes, probably owing to the temperature variations of the lens, ambient air, and telescope structure, and also with the change of the posture of the telescope during observation. The predicted change of the focus position during observation is calculated on the basis of an empirical formula, and the position of the camera is adjusted automatically.

The devices in the telescope such as the filter, the camera, and the focus adjustment mechanism are controlled by a Microsoft Windows personal computer (PC) in the observation control room. The microcomputer tuning the Lyot filter is also controlled by the PC. The cameras (and other devices in the telescope) and PCs are mostly connected with optical fibers to electrically isolate the devices from the PCs as much as possible.

The H$\alpha$ system captures images at several off-band wavelengths in addition to the H$\alpha$ center. These include H$\alpha\pm$0.05 nm and $\pm$0.08 nm to obtain Doppler signals and H$\alpha+0.35$ nm for context continuum images. The Doppler shift of 0.08 nm corresponds to the line-of-sight velocity of 37 kms$^{-1}$. Combining the structural changes seen in the images showing lateral motions and Doppler signals, the three-dimensional motions of cool plasma in the early phase of eruptive phenomena are derived. The line center images are recorded every 30 s, while the off-band images are taken every 150 s. 

The H$\alpha$ line center images are taken with three different exposure times, while the off-band images are taken with a single exposure time. The standard exposure for the H$\alpha$ center, 4 ms, is optimized for ordinary structures seen in H$\alpha$. In addition, images with a short exposure, 1 ms, which is adequate for bright kernels of flares, and those with a long exposure, 16 ms, for prominences outside the limb, are also captured. The long exposure time for the high SNR imaging of prominences became available after the installation of the sCMOS camera.

Images at each wavelength and each exposure time are obtained by taking 30 images consecutively. Thanks to the high frame rate of the camera, it takes less than 1 s. The best image is selected from the 30 images on the basis of the contrast of small-scale structures on the solar surface. Finally, only the best image is recorded.

A preliminary processing is applied to the obtained data in real time by another PC. A duty observer checks the quality of the images obtained. During this processing, the central wavelength of the transmission of the H$\alpha$ filter is also calculated. The transmission wavelength of the filter varies slightly with the ambient temperature. In case of a mismatch between the true transmission wavelength and the expected one, the reference point of the wavelength tuning is changed.

The calibration data, namely dark and flat images, are obtained using a calibration turret installed in the hood at the top of the telescope tube as shown in Figure 3. We adopted a trigonal pyramid for the shape of the turret to store the turret inside the hood, instead of the flat disk used usually. Three surfaces of the pyramid have an open aperture for usual data acquisition, a closed aperture for dark data, and an aperture with a diffuser for flat data. The diffuser is a light shaping diffuser (Edmund Optics) with a beam spreading angle of 1$^\circ$. The amount of diffused light entering the camera with a light shaping diffuser is much larger than that with a usual isotropic diffuser. A sCMOS camera has two A/D-conversion systems, one for low-light level and another for high-light level. Therefore, we take flat images with two exposure times, matching these two light levels.

Real-time processing and post-facto processing of the data produce quick-look pictures and movies as well as ``FITS'' (Flexible Image Transport System; see \url{https://fits.gsfc.nasa.gov/standard40/fits_standard40aa-le.pdf}) files using all of valid intensity and Doppler data. The images and Doppler data are acquired every 2.5 min; they are processed and uploaded to the webpage within 1 min after the data acquisition, namely in near real-time. The other data such as movies are uploaded daily. The real-time data have particular importance from the viewpoint of space weather, because they are expected to show ongoing flares and eruptive events, and three-dimensional motions of the erupted plasma can also be inferred.

\subsubsection{T4 Ca K, G-band, and continuum imaging system}

The Ca K, G-band, and continuum imaging system is installed in the rear of T4 as shown in Figure 3. The basic purpose of this system is to take context images of the photosphere and the chromosphere. It also has a 150 mm objective lens, and is used with a 125 mm aperture stop for compatibility with T1. The G-band and continuum images are obtained using the same camera, and the path to the Ca K camera and that to the G-band/continuum camera are divided by a beam splitter. 

The Ca K filter (Daystar) has a central wavelength of 393.4 nm and a bandwidth of 0.2 nm. 
The camera for the Ca K imaging is a Bitran BH-52L, formerly used to take H$\alpha$ images. The smear effect is not as severe as in the H$\alpha$ imaging, owing to the longer exposure time compared to the H$\alpha$. In addition, we do not intend to observe prominences in Ca K. 
As in T1, the predicted change of the focus position is automatically compensated. For Ca K, data is acquired every 30 s, and 10 images are taken at each data acquisition. The best image among them is recorded.

In the G-band/continuum system, a filter turret is used to select the wavelength. This contains an interference filter for the G-band and a green broad-band filter for the continuum. Their central wavelengths and the bandwidths are 430.5 nm / 1.0 nm and 530 nm / 50 nm, respectively.
To compensate the difference of the focus positions between the continuum and the G-band, a glass block is added to the continuum path.

A Photonfocus MV1-D2080-160 CMOS camera is used for the G-band/continuum imaging. In spite of its noise level being not very low, its full-well capacity (about 90k e$^-$), which is large for a high-speed camera, realizes a fairly high SNR for signals close to the full-well level. For the G-band/continuum, data are acquired every five minutes, and 30 images are captured for each wavelength at each data acquisition. The best quality image among them is recorded.

These three wavelengths, which are considerably shorter than the H$\alpha$, undergo atmospheric absorption more severely than the H$\alpha$, when the Sun is at low elevations. Therefore, we adjust the exposure time according to the brightness of the solar disk at each wavelength in the range of 1$\times$ -- 4$\times$ of the standard exposure times to ensure that the signal level around the disk center does not become much lower than the full-well level.

A calibration turret, the same as that of T1, is also installed in the hood at the top of the telescope tube. The dark and flat images are taken using the calibration turret. 

The obtained data are processed after daily observation. A set of images is extracted every hour, and their quick-look pictures of Ca K, G-band, and continuum are uploaded to the webpage. 

\subsection{Infrared spectropolarimeter}

\begin{figure}
\centering
\includegraphics[width=0.7\columnwidth]{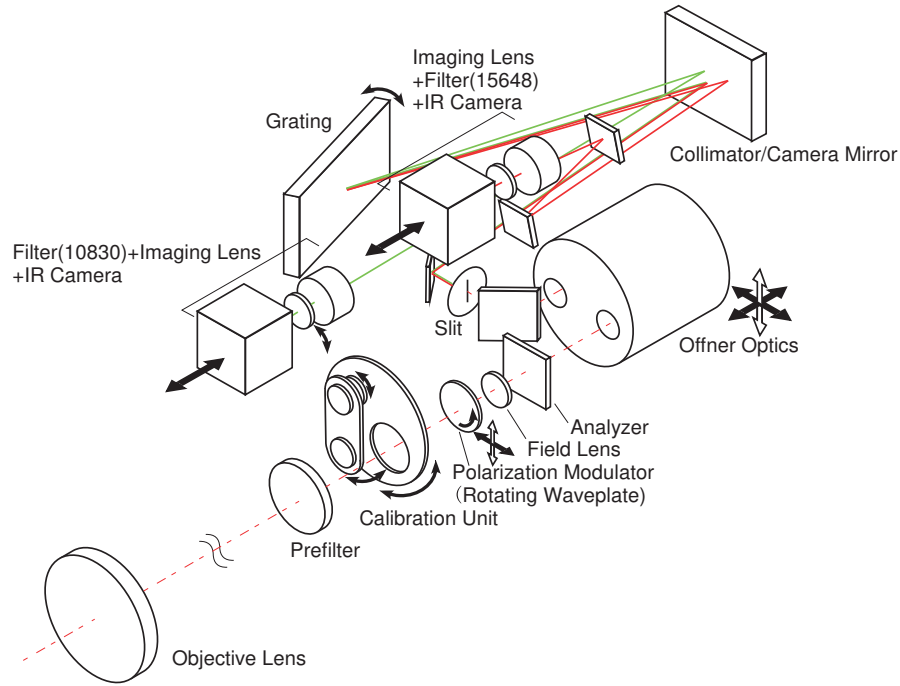}
\caption{\small Schematic diagram of the infrared spectropolarimeter at the rear of T2 (and T3). Red dash-dot lines show the optical axis before the spectrograph slit, while green and red solid lines represent the optical axes of the 1083.0 nm and the 1564.8 nm light in the spectrograph, respectively. The arrows indicate the motion of movable parts.
} 
\label{fig:fig4}
\end{figure}

Since the infrared spectropolarimeter has been described in \citet{2018PASJ...70...58S} in detail, we present a brief overview of the instrument here.

A schematic diagram of the infrared spectropolarimeter is shown in Figure 4. The objective lens fabricated by Genesia, who also assembled the spectrograph system, has a diameter of 150 mm, and it has the same focus for the two wavelength ranges, 1083.0/1082.7 nm and 1564.8 nm. An interference filter fabricated by Optical Coatings Japan, which transmits only these wavelength ranges, is placed just downstream from the objective lens to reject unnecessary light. A polarization modulator is installed before the folding mirrors to prevent generating instrumental polarization. The polarization modulator was two ferroelectric liquid crystals until 2013 August, when it was replaced by a rotating waveplate. A polarizer is placed after the polarization modulator as an analyzer; this is a single-beam polarimeter using only one of the orthogonal linear polarizations. 

Unlike the filter imaging instruments described in the previous subsection, this is a spectrometer. Therefore, slit scan is necessary to cover the full Sun. On the other hand, the telescope needs to be always pointed at the disk center for the imaging observations. We have an Offner optics (consisting of two spherical mirrors) before the spectrometer slit to move the solar image on the slit and cope with the two types of observations.
The rotating waveplate unit is correspondingly moved with the scan with the Offner optics to ensure that the light arriving at the slit always passes through the center of the rotating waveplate. The Offner optics also serves as focus adjustment optics.

The two cameras (XEVA640 of XENICS with InGaAs detectors) cover the wavelength ranges of 1083.0/1082.7 nm and 1564.8 nm, respectively. The spatial sampling on the cameras is matched with the diffraction limit of the objective lens of about 2--3$''$. Because they are small format cameras (640$\times$512 pixels), the detectors cover only about half of the solar diameter along the slit. The slit moves stepwise every about 2$''$, and separate scans are performed for the northern and southern hemispheres to cover the full-disk. Currently, at a slit position, 192 images (8 images $\times$24 modulations) are taken during 12 rotations of the waveplate of the modulator, and it takes about 3 s. These images are integrated and converted to Stokes data. A typical noise level of the Stokes data was $5\times 10^{-4}$ until 2013, but from 2014, it was reduced to about $1.6\times 10^{-4}$ because of the improvement of the efficiency of the polarization modulation brought by the replacement of the polarization modulator. A complete scan requires about 2 hours. While observations started with a single camera, a second one was installed in 2015 to record the spectra of the two wavelength ranges simultaneously. At present, a couple of sets of full-disk data of the two wavelength ranges are being collected daily.

As in T1 and T4, the focus positions of the image on the slit and those of the spectra on the camera change during observation. The predicted changes of the focus positions are automatically compensated.

During a full-disk scan for about two hours, clouds may pass across the Sun. When a cloud is detected by at least one of the two solar brightness and position sensor systems (one for the guiding system of the telescope and another for the slit monitor consisting of four linear sensors), the slit movement stops and data acquisition is resumed when data without cloud are obtained. This avoids data loss due to cloud passage.  

The obtained data are processed after daily observation, and quick-look pictures for all the scans are uploaded to the web-page.
Currently the circular polarization maps of 1083.0/1082.7 nm and 1564.8 nm as well as the intensity maps of 1083.0 nm are presented in the webpage. In addition, we are now preparing to replace them with magnetograms.

\section{Significance of the SFT observation in space weather research}

At present, observations with the SFT can obtain data for chromospheric dynamic phenomena such as flares and filament eruptions as well as the magnetic field which drives these dynamic phenomena.
This combination has a particular advantage from the point of view of space weather. Here we present some examples demonstrating the significance of such observations \citep[see also,][]{2019BAAS...51c.110M}.

\subsection{An event on August 6--8, 2012}

\begin{figure}
\centering
\includegraphics[width=1.0\columnwidth]{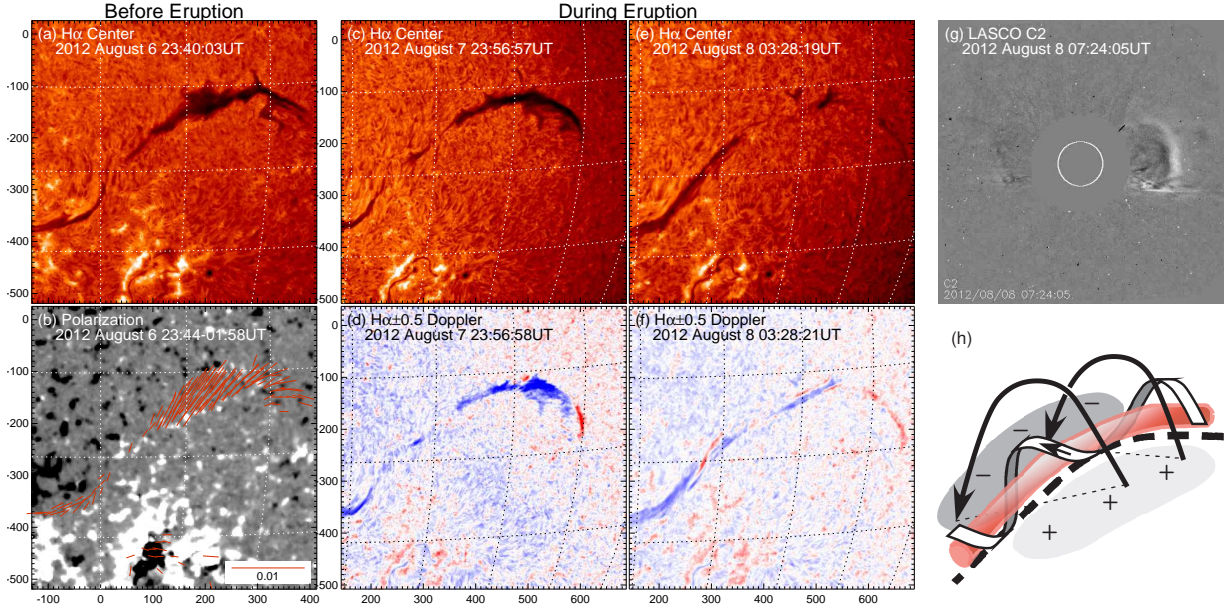}
\caption{\small Filament eruption and accompanied CME occurring during August 6--8, 2012. (a)(b) A H$\alpha$ image and a polarization map including the filament before the eruption taken with the SFT. 
Red lines in panel (b) show the linear polarization of He I 1083.0 nm; the length of each line indicates the degree of the corresponding linear polarization signal (the line in the inset corresponds to the polarization of 0.01 [1 \%]), and the its direction indicates the orientation of the polarization signal.
Only the linear polarization signals $>$0.1\% are shown. The background gray-scape map shows the circular polarization of Si I 1082.7 nm, which provides the photospheric longitudinal magnetic field; the display range is $\pm 1$ \%.
 (c)--(f) The filament during its eruption observed with the SFT. Panels (c) and (e) show H$\alpha$ line center images, and (d) and (f) show the Dopplergrams based on the H$\alpha\pm$0.05 nm images. Blue and red shifts of the H$\alpha$ line are indicated by the corresponding colors. (g) A CME seen several hours after the filament eruption observed with the LASCO C2 of SOHO, of which the field of view covers up to 6 solar radii. (h) Schematic representation of the magnetic field structure around the filament before its eruption. The red bar represents the filament material, and helical magnetic field of the flux rope surrounds it. Coronal magnetic loops connect large scale positive and negative polarities on the photosphere.
} 
\label{fig:fig5}
\end{figure}

Figure 5 shows a filament and its eruption, which developed into a CME, observed during August 6--8, 2012. The filament and its magnetic field before the eruption, as well as the ongoing eruption, were observed with the SFT. A day before the eruption, a stable quiescent filament was seen in a H$\alpha$ picture (Fig. 5a). In Figure 5b, the linear polarization signals of He I 1083.0 nm with a linear polarization degree $>$0.1\% are indicated by the red lines on a circular polarization map of Si I 1082.7 nm showing the photospheric magnetic field. The length of each red line represents the degree of polarization of the corresponding linear polarization signal and the direction represents the orientation of the polarization signal determined by $\frac{1}{2}\arctan ({\rm Stokes}~U/Q)$. The filament is located at the boundary of positive (white in the circular polarization map) and negative (black) polarity areas in the photosphere.
Figure 5b indicates that conspicuous linear polarization signals of He I 1083.0 nm well above the noise level (about $5\times 10^{-4}$) are concentrated in the filament.

The linear polarization seen in the filament is produced by atomic level polarization \citep{2002Natur.415..403T} and not by Zeeman effect, which is the most common source of polarization in the solar atmosphere. 
Atomic level polarization is produced particularly in a plasma cloud in the corona such as filaments, since they receive anisotropic radiation arising only from below. Atomic level polarization produces distinct net polarization. The orientation of the linear polarization signals in the filament is parallel to the magnetic field line, and therefore, the red lines in Figure 5b are considered to display the distribution of the transverse magnetic field in the filament. 
The linear polarization signals deviate from the filament axis similarly throughout the filament, and the deviation is counterclockwise with respect to the filament axis. This is a common property of filaments in the southern hemisphere \citep{2017ApJ...851..130H}. Figure 5h shows a schematic picture of the magnetic field structure around the filament. In the corona, overlying magnetic field above the polarity inversion line is usually right-skewed in the southern hemisphere \citep[e.g.,][]{1998ASPC..150..419M}. A helical flux rope is presumed to be located below the coronal loops. The observed polarization signals in Figure 5b are considered to be produced by the magnetic field at the bottom of the flux rope, where the cool material of the filament is piled up.

A day later, the filament began to erupt as shown in Figures 5c and 5d. While the filament is still visible in the image taken at the H$\alpha$ center (Fig. 5c), most of the filament shows conspicuous blue shift in a Dopplergram (a map of Doppler signals), produced from the H$\alpha\pm$0.05 nm images (Fig. 5d). This indicates that the filament has already started to move upward at this moment, and the line-of-sight velocity of the most remarkable part is estimated to be about 20 km s$^{-1}$ toward the observer. Unfortunately, polarimetric data could not be obtained on the day of the eruption owing to frequent cloud passage. About 3.5 hours later, as seen in in Figure 5e, the most prominent part of the filament in Figure 5c erupted away. About four more hours later, a CME was seen in a running difference image captured using the Large Angle Spectrometric Coronagraph \citep[LASCO;][]{1995SoPh..162..357B} C2 on board the Solar and Heliospheric Observatory \citep[SOHO;][]{1995SoPh..162....1D} shown in Figure 5g. The magnetic structure around the filament shown in Figure 5h is expected to develop into an interplanetary magnetic flux rope.

As shown above, the H$\alpha$ data in this event, including the Doppler velocity measurements, reveal the dynamic behavior of the erupting filament, while the He I 1083.0 nm data show the magnetic field in the filament before its eruption. 
This fact means that, from the viewpoint of space weather, the combination of the high-cadence imaging observation and the polarimetric observation provides information about the dynamics and the magnetic structure of a flux rope before and during the early phase of its eruption. The combination of these two types of observations are essential for the synoptic telescopes monitoring the Sun in the present day and future. The SFT enabled it.

\subsection{An event on July 9, 2013}

\begin{figure}
\centering
\includegraphics[width=0.7\columnwidth]{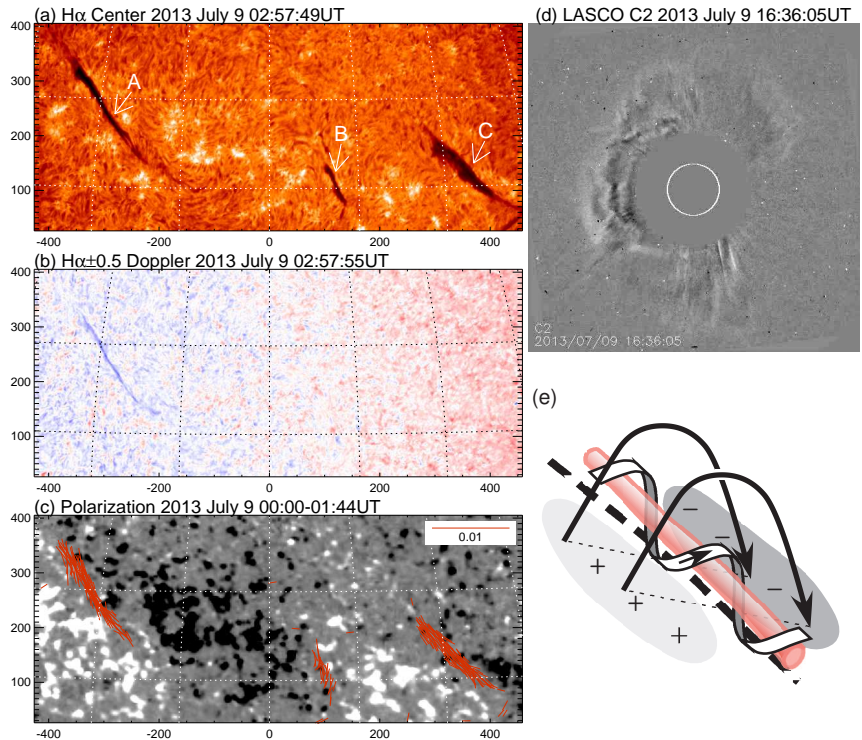}
\caption{\small Filament observed before its eruption and CME occurring on July 9, 2013. (a)(b) The filament before its eruption observed with the SFT. Panel (a) shows a H$\alpha$ line center image, and (b) shows a Dopplergam based on the H$\alpha\pm$0.05 nm images. Blue and red shifts of the H$\alpha$ line are shown in corresponding colors. Three filaments are seen in panel (a), where the one on the left labelled ``A'' erupted later. (c) A polarization map including the filament before eruption taken with the SFT. 
Red lines show the linear polarization signals of He I 1083.0 nm; the length of each line indicates the degree of the corresponding linear polarization signal (the line in the inset corresponds to the polarization of 0.01 [1 \%]), and the its direction indicates the orientation of the polarization signal.
Only the linear polarization signals $>$0.1\% are shown. The background gray-scape map shows the circular polarization of Si I 1082.7 nm; the display range is $\pm 1$ \%.
(d) A halo CME seen after the filament eruption observed using the LASCO C2 of SOHO, of which the field of view covers up to 6 solar radii. (h) Schematic representation of the magnetic field structure around the erupted filament. The red bar represents the filament material, and helical magnetic field of the flux rope surrounds it. Coronal magnetic loops connect large scale positive and negative polarities on the photosphere.
} 
\label{fig:fig6}
\end{figure}

The erupted plasma in Figure 5 did not reach the Earth. However, there is another example of the magnetic field observation of a filament, which erupted later and developed into a geo-effective interplanetary CME (ICME). 
Figures 6a and 6b show the filament observed in H$\alpha$ taken with the SFT before its eruption on July 9, 2013. Three filaments are found in the quiet region in Figure 6a designated by arrows. The left filament (labelled ``A'') began to erupt around 14 h on the same day, about half a day later, according to the 30.4 nm images taken with the Atmospheric Imaging Assembly (AIA) of NASA's Solar Dynamics Observatory (SDO). In a Dopplergram (Fig. 6b), a weak blue shift, which corresponds to the line-of-sight velocity of 5--10 km s$^{-1}$ toward the observer, can be found in the filament that erupted later, while the other two labelled ``B'' and ``C'' show no Doppler signals. This indicates that some kind of activation of the filament was already underway at this moment, though it was half a day before the actual eruption. After the observation of Figures 6(a)--(c), the filament erupted. Since it was in the nighttime in Japan, the SFT did not observe the eruption. It developed into a halo CME as shown in a running difference image captured by LASCO of SOHO (Fig. 6d). The eruption developed into an ICME, which hit the Earth \citep{2018SpWea..16..216M}, causing a moderate geomagnetic storm. 

Figure 6c shows the linear polarization signals of the He I 1083.0 nm with red lines drawn on a circular polarization map of Si I 1082.7 nm also taken with the SFT. Again, most of the linear polarization signals are concentrated in the filaments. The magnetic field of the filament which erupted later shows the ordinary chirality according to the hemispheric rule \citep{1998ASPC..150..419M} as well as the other two stable filaments. Figure 6e shows a schematic picture of the magnetic field structure around the filament. The magnetic field structure of a filament which is about to erupt can be predicted using the polarization data.

According to \citet{2018SpWea..16..216M}, in addition to the various spacecrafts dedicated to solar observation, MErcury Surface, Space ENvironment, GEochemistry and Ranging (MESSENGER) of NASA, which was located near the Sun-Earth line, were available at the moment of the CME. Therefore, particularly many-sided observations of the CME and the flux rope, both coronagraphic and {\it in situ,} were performed. These data formed the basis of simulations of the development of the flux rope which reproduced its geomagnetic effect. This is one of the attempts at predicting the southward magnetic component of flux ropes, which drives geomagnetic storms, and studies such as these are attracting more attention from the point of view of operational space weather programs.

In such attempts, the magnetic field in the source region of CMEs is crucial. Many studies have been undertaken to investigate the relationship between the magnetic field in the source region of CMEs and the interplanetary flux ropes \citep[e.g.,][]{2014ApJ...793...53H, 2015SoPh..290.1371M, 2017SoPh..292...39P, 2018SpWea..16..442P}. These studies have inferred the magnetic field structure of the source region of ICMEs from various features on the Sun such as the H$\alpha$ fine structures in filaments, shape of the overlying coronal loops, and photospheric magnetic field information. However, they are indirect information. 
The magnetic field properties of ICME flux ropes, such as field strengths and twists, are difficult to determine using only indirect information. In particular, the axial twist of the magnetic field is quite uncertain and is difficult to determine correctly \citep[see e.g.,][]{2014ApJ...793...53H}. Utilizing the polarization data obtained from the observations of the He I 1083.0 nm, which show the magnetic field in filaments, are expected to aid such studies, even if their availability is limited to cases where the CMEs are accompanied by filament eruptions. Actually, \citet{2020ApJ...892...75W} and \citet{2020arXiv200610473K} successfully carried out the detailed analysis of the magnetic field in filaments before and during their eruption. The observation of the SFT is expected to provide data for such analyses routinely.

\section{Concluding remarks}

The solar group of NAOJ is conducting synoptic solar observation including imaging at H$\alpha$, Ca K, G-band, and continuum, and spectropolarimetry at wavelength bands including the He I 1083.0 nm / Si I 1082.7 nm and the Fe I 1564.8 nm lines using the SFT. These observations measure the brightness distribution, Doppler signal, and magnetic field information of the photosphere and chromosphere. 
As shown in the examples of observational data in Section 3, the H$\alpha$ imaging and Doppler observations enable the estimation of three-dimensional velocities of eruptive features in the early phase of CMEs. Polarization measurements at He I 1083.0 nm show the magnetic field structure in filaments, which develops into CME flux ropes in the interplanetary space.
CME flux ropes occasionally hit the Earth. 
Hence, data such as those taken with the SFT synoptic observation are expected to contribute to space weather research facilitating the prediction of geo-effective ICMEs in the early phase of the events.
Such an ability to contribute to space weather research is the advantage of the SFT, which most of the previous synoptic instruments does not have.

This explains the proposals for further advanced synoptic instruments focusing on space weather research, which include infrared spectropolarimety like the SFT. Current synoptic observations with the SFT are expected to become a pathfinder for future advanced synoptic instruments, in spite of certain limitations in these observations. These limitations include the fact that these observations are being carried out at a single station with a rather small telescope, with a rather low cadence of polarization data acquisition. 
We are developing a large format infrared camera to ensure efficient data acquisition with a higher cadence for future observations \citep{Hanaoka2019}.

NAOJ has been conducting solar monitor observations for more than 100 years. The current observation, which basically focuses on short-term solar activity, also aims at continuing long-term activity monitoring. For instance, the Ca K data currently obtained using the SFT are used to reconstruct the long-term variation of the solar activity as a part of the worldwide accumulation of Ca K data as well as historical photographic spectroheliograms taken at the Tokyo Astronomical Observatory, one of the predecessors of NAOJ \citep{2020A&A...639A..88C}.
Long and continuous observations are a key to elucidate the dynamo activity of the Sun, and therefore, long-term observations are also necessary to study the solar activity and its influence on the Earth.
The observations of solar activity phenomena carried out using state-of-the-art instruments facilitate the long-term acquisition of activity data.

\begin{acknowledgements}
This work is supported by JSPS KAKENHI Grant Number JP17204014, JP19540244, JP23244035, and JP15H05814, and also by the NAOJ research grant. The instruments were developed and tested with the support of the Advanced Technology Center of NAOJ.
We appreciate the staff members of the Solar Science Observatory who have been involved in the daily operation of the instruments and maintenance of the data servers. 
The SOHO/LASCO data used
here are produced by a consortium of the Naval Research
Laboratory (USA), Max-Planck-Institut f\"{u}r Aeronomie (Germany),
Laboratoire d'Astronomie Spatiale (France), and the
University of Birmingham (UK). SOHO is a project of
international cooperation between ESA and NASA.
We would like to thank Editage (www.editage.com) for English language editing.
The authors are grateful to the anonymous referees and the editor for their careful reading and helpful comments.
\end{acknowledgements}

\bibliography{hanaoka_arXiv}
   

\end{document}